\documentclass[aps,prd,twocolumn, preprint]{revtex4-2}
\usepackage{graphicx}
\usepackage{bm}
\usepackage{soul, color}

\newcommand{\ColFigWidth}{.5\textwidth}

\begin{document}

\title{Zonal fields as catalysts and inhibitors of turbulence-driven magnetic islands}% Force breaks with \\

\author{D. Villa}
 \affiliation{Max Planck Institute for Plasma Physics, Boltzmannstraße 2, 85748, Garching, Germany}
 \affiliation{Excellence Cluster ORIGINS, Boltzmannstraße 2, D-85748, Garching, Germany}
 \email{daniele.villa@ipp.mpg.de}

\author{N. Dubuit}%
\affiliation{Aix-Marseille Universit\'e, CNRS, PIIM UMR 7345, Marseille, France}

\author{O. Agullo}
\affiliation{Aix-Marseille Universit\'e, CNRS, PIIM UMR 7345, Marseille, France}

\author{X. Garbet}
\affiliation{CEA, IRFM, F-13108 Saint-Paul-Lez-Durance, France}
\affiliation{School of Physical and Mathematical Sciences, Nanyang Technological University, 637371 Singapore}

\date{\today}% It is always \today, today,
             %  but any date may be explicitly specified

\begin{abstract}
A novel coalescence process is shown to take place in plasma fluid simulations, leading to the formation of large-scale magnetic islands that become dynamically important in the system. The parametric dependence of the process on the plasma $\beta$ and the background magnetic shear is studied, and the process is broken down at a fundamental level, allowing to clearly identify its causes and dynamics.
The formation of magnetic-island-like structures at the spatial scale of the unstable modes is observed quite early in the non-linear phase of the simulation for most cases studied, as the unstable modes change their structure from interchange-like to tearing-like. This is followed by a slow coalescence process that evolves these magnetic structures towards larger and larger scales, adding to the large-scale tearing-like modes that already form by direct coupling of neighbouring unstable modes, but remain sub-dominant without the contribution from the smaller scales through coalescence. The presence of the cubic non-linearities retained in the model is essential in the dynamics of this process. The zonal fields are key actors of the overall process, acting as mediators between the competitive mechanisms from which Turbulence Driven Magnetic Islands can develop. The zonal current is found to slow down the formation of large-scale magnetic islands, acting as an inhibitor, while the zonal flow is needed to allow the system to transfer energy to the larger scales, acting as a catalyst for the island formation process.
\end{abstract}

%Use showkeys class option if keyword
%display desired
\maketitle

% Word count before modifications (i.e. word count of "PRL_Resubmit" = 3692)

% \section{Introduction}

\begin{figure}
\centering
\includegraphics[width=.48\textwidth]{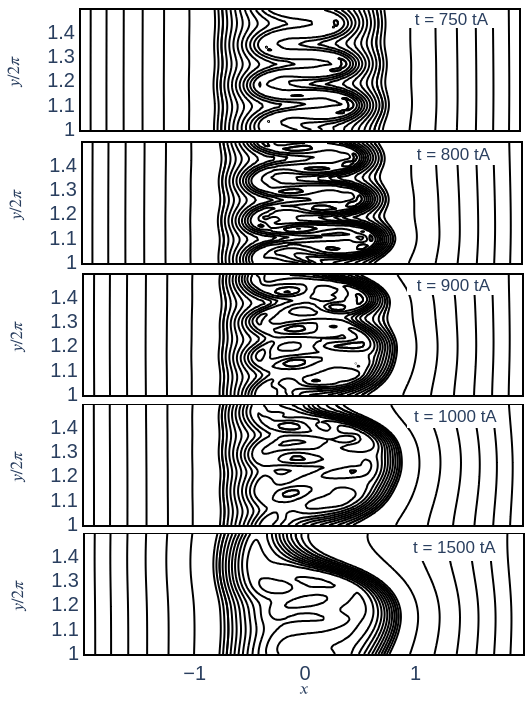}
\caption{\label{fig_coal_seq}Sequence of isocontours of $\psi$ taken from the simulation with $\beta = 1.28 \%$ and $\partial_x B_{eq} = 0.02$ showing the coalescence process taking place from the late linear phase (top) to the non-linear phase (bottom).}
\end{figure}

Magnetic islands are a well-studied phenomenon in plasma physics, being the product of magnetic reconnection, a process that leads to the re-organization of the magnetic configuration and lowers the overall internal energy of the system. Magnetic reconnection itself is a phenomenon of interest to the whole domain of magnetized plasma physics, as it leads to localized heating and changes in the magnetic field topology, thus strongly affecting the properties and dynamics of the plasma \citep{falconer2000assessment, barta2008dynamics, janvier2017three}. In case the system allows for it, the newly formed magnetic structures can form subsets of nested flux surfaces that can act as a localized confinement region, the magnetic island itself.\\
Despite the large volume of work done on magnetic islands \citep{waelbroeck2009theory, poye2014dynamics, ishizawa2019multi, choi2021interaction, betar2022microscopic}, there are still open questions about the fundamental properties of these structures and the dynamics of reconnection, namely how exactly turbulence drives reconnection, thus forming Turbulence Driven Magnetic Islands (TDMIs), what are the differences in the dynamics for a plasma that undergoes reconnection at different plasma $\beta = \frac{2\mu_0 \; p}{B^2}$ regimes (where $p$ is the plasma pressure, $B$ the magnetic field strength and $\mu_0$ the vacuum permeability constant) and how do zonal fields (i.e. axisymmetric flows and currents) interact with magnetic islands. Of particular interest for this work are the dynamics of turbulent reconnection \citep{lazarian1999reconnection, ishizawa2010turbulence, lazarian20203d, choi2021interaction}, the generation of magnetic islands by turbulence \citep{yagi2007nonlinear, muraglia2011generation, agullo2017nonlinearI, dubuit2021turbislands} and the interaction of magnetic islands with zonal fields \citep{guzdar2001zonal,diamond2005zonal,fujisawa2007experimental,dong2019nonlinear}.\\
In this letter, the generation of magnetic islands by turbulence is shown in systems with varying plasma $\beta$, magnetic shear and instability drive. The relation of the new process to those previously explored in the literature \citep{yagi2007nonlinear, muraglia2011generation} is addressed, and, as will be detailed, this new process comes in addition to, and supercedes, previous processes leading to TDMI formation, in particular in near-marginal regimes relevant to low-$\beta$ plasmas. The core finding of this paper is illustrated in Fig. \ref{fig_coal_seq}, where the transition of the system from the linear interchange instability to the non-linear formation of TDMIs through coalescence is shown as a sequence of isocontours of the field $\psi$ at different times.
% Removed 70 words

% \section{Non-linear simulations: an electromagnetic 6-field reduced 2-fluid model} \label{sec_model}

To demonstrate the presence of these mechanisms and assess their importance, a reduced 6-field fluid model is considered, developed starting from the Braginskii fluid equations \cite{braginskii1965transport, villa2022localized}, allowing to describe both small-scale phenomena, like interchange instabilities and the deriving turbulence, and large-scale ones, like magnetic islands and zonal fields, on the time scales needed to study (resistive) magnetic reconnection. The model evolves 6 dynamic fields: the magnetic potential $\psi$, the electrostatic potential $\phi$, the electron and ion pressure $p_{e/i}$, the density $n$ and the ionic parallel velocity $u_\parallel$. Reduction of the system to single-helicity ``2D'' slab geometry allows studying the fundamentals of the process in a simplified and generic environment, where the only requirement is a strong background magnetic field that allows defining the geometry of the system in reference to a chosen magnetic helicity (that identifies a ``resonant position'' \citep{waelbroeck2009theory, Scott1992Long}). The background magnetic field is used to define a parallel and a perpendicular direction and establish drift-ordering expansions \citep{scott2001low2} (that are used to derive the model), that allow to study slow processes $\omega / \Omega_i \ll 1$, where $\Omega_i$ is the ion cyclotron frequency and $\omega$ the characteristic timescale of the studied phenomenon. The model being a ``reduced'' model means that fluctuations of the magnetic field in the parallel direction are considered to be higher order corrections to the dynamics, i.e. $\tilde{B}_\parallel/B_0 \ll \tilde{B}_\perp/B_0 < 1$. The $\sim$ indicates fluctuations, the parallel ``$\parallel$'' and perpendicular ``$\perp$'' directions are defined relative to the total magnetic field, and $\partial_z B_z = 0$, with $z$ being the direction of the dominant component of the background magnetic field. Thus (after normalization) one can write the fluctuations of the magnetic field as $\bm{\tilde{B}} = \nabla \times (\psi \bm{\hat{z}}) $. In slab geometry, the parallel derivative of a generic function $f$ can be expressed in a convenient form keeping into account the fluctuations of $\bm{\tilde{B}} = \bm{\tilde{B}_\perp}$ through the Poisson brackets: $\bm{\tilde{B}} \cdot \nabla f = \{\psi, f \} = (\nabla \psi \times \nabla f ) \cdot \bm{e}_{z}$ where $\nabla$ is the gradient and $\bm{e}_z$ the direction of the background out-of-plane magnetic field. The explicit expression for the Poisson bracket in slab geometry is: $ \lbrace \psi, f \rbrace = ( \partial_x \psi \, \partial_y f - \partial_x f\, \partial_y \psi )$. Here, $x$ is the radial direction, $y$ the periodic poloidal direction and any field $f$  can be decomposed according to $f=\sum\limits_{m = -\infty}^{+\infty}f_m(x) \exp(2i\pi m y/L_y + \varphi)$.\\
The full expression of the flux divergence terms is retained in the model, meaning that, in reduced notation, the parallel derivatives are given by the product of $3$ fluctuating fields, e.g. $ \nabla_\parallel (p u) = p \nabla_{\parallel} u_{\parallel} + u_{\parallel} \nabla_\parallel p = p  \lbrace \psi , u_{\parallel} \rbrace + u_{\parallel} \lbrace \psi , p \rbrace$. Such terms, also used in other models \citep{giacomin2021gbs, pitzal2023landau}, are referred to as ``cubic terms'' \citep{villa2022cubic}, and it has been shown \citep{villa2022localized} that they can significantly impact the dynamics. Velocities in the system are normalized to the Alfvén velocity at the reference position $v_A = \frac{B_0}{\sqrt{\mu_0 m n_0}}$, lengths are normalized to a characteristic but arbitrary perpendicular length scale $L_\perp$, the magnetic field is normalized to the reference value $B_0$ and both pressure and density are also normalized to the reference values at the resonant magnetic surface. A more detailed description of the model, including simulation parameters, is provided in the supplementary material and/or in the references \citep{villa2022localized}. This model is implemented in the fluid code AMON \citep{poye2012dynamique}.\\
For the simulations of this work, only the interchange instability is present, driven by the coupling of the equilibrium ion pressure gradient to the curvature of the magnetic field. Having a single-helicity system implies that there is only one resonant position, where the magnetic field determines the reference helicity, which allows for clearer analysis of the mechanism. To limit the number of free parameters, the only background gradients in the system are those of the ion pressure and of the magnetic field. Note that in order to have a net 0 perpendicular equilibrium drift one has to impose $\partial_x \phi_{eq} = - \Omega_i \tau_A \rho^2_* \; \frac{\partial_x p_{i \, eq}}{n_{eq}}$. The magnetic field is linearly stable to the tearing mode, which is confirmed by computing, at the resonant position $x = 0$ the parameter $\Delta' \leq -1.9$ \citep{furth1963finite, arcis2006rigorous}. This value of $\Delta'$ is obtained by using a Harris-sheet background magnetic field with sufficiently weak gradient. The resistivity $\eta = 10^{-5}$ implies a typical time-scale for resistive reconnection $\tau_\eta = L_\perp^2/\eta \approx 10^4 \; \tau_A $ for a typical $L_\perp = 0.5$, meaning that phenomena occurring on time-scales faster than $\tau_\eta$ are due to turbulent dynamics. The simulations are thus gradient-driven, though the mode $m=0$ is self-consistently evolved, allowing for saturation dynamics to take place, should they be present.

% Removed -40 words (i.e. added)

% \section{Generation of magnetic islands by interchange turbulence} \label{sec_generation}

\begin{figure}
\includegraphics[width = \ColFigWidth, trim=0 0 0 2.75cm, clip]{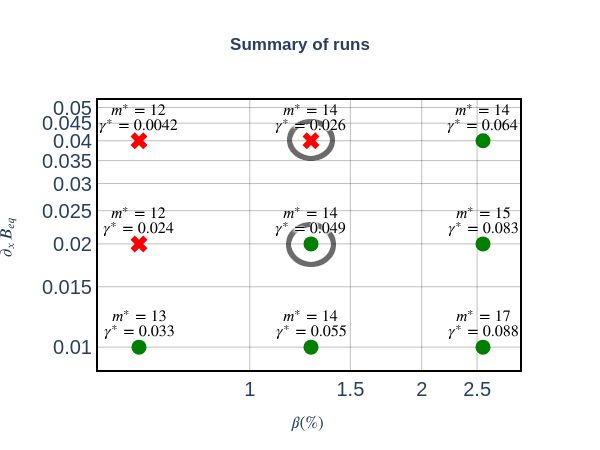}
\caption{\label{fig_PRL_summary} Summary of results for 2D slab non-linear runs of linearly unstable interchange where the plasma $\beta$ and background magnetic shear were varied. Green circles represent simulations that showed the formation of magnetic islands and at least the initial phases of coalescence, while simulations indicated with red ``X''s didn't. The mode number $m^*$ and the corresponding growth rate $\gamma^*$ on top of the symbols refer to the most unstable mode. Circled simulations are used in Fig. \ref{fig_NL_eigenmode}}
\end{figure}

\begin{figure}
\centering
\includegraphics[width=\ColFigWidth, trim=0 0 0 2.75cm, clip]{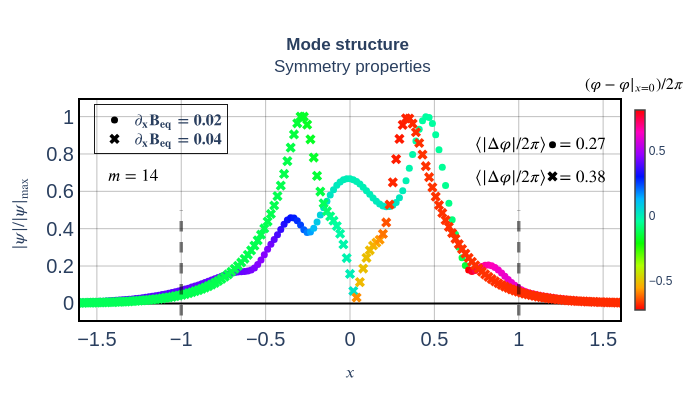}
\caption{\label{fig_NL_eigenmode}Radial structure and phase differences of the mode $m=14$ of $\psi$ for two non-linear simulations with $\beta = 1.28 \%$ but different magnetic shear as indicated in the legend ($\partial_x B_{eq} = 0.02$ forms TDMIs, $\partial_x B_{eq} = 0.04$ doesn't). The differences in phase with the reference value at the resonance are computed across the radius, along with the average $\langle | \Delta \varphi | / 2 \pi \rangle$, over the portion of the radius between the two dashed vertical lines at $x = \pm 1$. Notice that for $\partial_x B_{eq} = 0.02$ the mode has more even (tearing-like) parity. A noticeable lower average phase difference shows up consistently through simulations that form TDMIs for all modes with $m < m^*$.}
\end{figure}

Non-linear simulations are run scanning the parameter space as shown in Fig. \ref{fig_PRL_summary}. All simulations are interchange unstable. This identifies a threshold in $\beta$ and $s = \frac{\partial_x B_{eq}\vert_{x=0}}{B_{eq}\vert_{x=0}} = \partial_x B_{eq}\vert_{x=0}$, the magnetic shear, for the formation of TDMIs. The definition of the magnetic shear varies in the literature, here $B_{eq}$ is the total equilibrium magnetic field, including the constant out-of-plane component, that appears at the denominator when writing $B_y$ as a function of the total magnetic field in a mono-helicity system, hence $\vert B_{eq}\vert_{x=0} \vert = 1$ and $s$ measures the inverse of the length scale of the variation of the ``poloidal'' magnetic field, but results presented in this paper are independent of the definition of the magnetic shear. Additional simulations have shown that the threshold is not related to the growth rate of the most unstable mode, but rather to the physical mechanisms acting at different shears, since simulations with very similar growth rates of the most unstable mode did or didn't show the formation of TDMIs depending on magnetic shear.\\
Notice that the weaker the magnetic shear, the more likely the appearance of TDMIs, reinforcing the point that the structures presented here do not originate from tearing \citep{furth1963finite}. The non-linear dynamics of the system further weaken the magnetic shear, and non-linear de-stabilization of tearing can also be ruled out.\\
What is actually meant by ``formation of TDMIs'' is illustrated in Fig. \ref{fig_NL_eigenmode}, where the radial structure of the mode $m=14$ for two simulations with $\beta = 1.28 \%$ but different magnetic shears $\partial_x B_{eq} = 0.02$ and $\partial_x B_{eq} = 0.04$ are compared. Since the modes are complex valued, the plot uses the absolute value of the function to represent its magnitude, and the colour scale to represent the phase difference of the mode at that particular position to the phase at the resonant position. For a mode with interchange parity, in a linear simulation, the eigenfunction of $\psi$ is odd (meaning that the real and imaginary parts are odd), corresponding to an even absolute value with a change in phase across the resonance, while for tearing parity the phase is constant across the resonance, corresponding to an even eigenfunction. Since simulations are non-linear, the words ``even'' and ``odd' are not to be understood as exact properties, but refer to the ``dominant'' component of the function, which does not necessarily match the linear one. Focusing on $\psi$, as the formation of magnetic islands is directly visible through this field, Fig. \ref{fig_NL_eigenmode} refers to $t = 1700 \tau_A$ (i.e. early in the non-linear phase) in two simulations where, at that particular time, the mode $m=14$ had the highest energy: the simulation that develops TDMIs ( $\partial_x B_{eq} = 0.02$ ) has a broad region of weakly varying phase around the resonance, whereas the simulation that doesn't form TDMIs has a clear variation in phase across the resonance. Averaging the absolute values of these phase differences $\Delta \varphi = (\varphi - \varphi_{res})/(2\pi)$ for $-1 \leq x \leq 1$, where the amplitude $A$ of the modes satisfies $A/A_{\max} \geq 0.1$, gives the average $\langle \vert \Delta \varphi (x) \vert \rangle_{x \in [-1,+1]}$ values in Fig. \ref{fig_NL_eigenmode}, showing a noticeable lower average phase difference when TDMIs form. Using the phase information to quantify the parity of the modes also allows to compensate for the shift of the resonant position in the non-linear phase, that would act as the symmetry center to define the parity of the mode. Performing this analysis over time and for all modes shows this behaviour of the average phase difference to be consistent over time, at least for all modes with mode-number $m \lesssim m^*$ ($m^*$ being the mode-number of the linearly most unstable mode). Thus, in simulations where the formation of TDMIs is observed, modes develop, on average, more tearing-like structures in their non-linear evolution, even though they start out with interchange parity.\\
Whether this change in structure of the unstable modes takes place determines whether or not the large-scale magnetic islands will form. What allows this change in structure is a combination of factors, including the velocity shear, whose role, along with that of the zonal current, will be expanded upon in the following section. However, suppressing the fluctuations (and the equilibrium) of the mode $\phi_0$ does not prevent the change in parity, so the mechanism requires also the parameter dependencies illustrated in Fig. \ref{fig_PRL_summary} and the cubic non-linearities retained in the model. This latter point has been verified by running simulations without cubic non-linearities, that don't show change in parity of the unstable modes. Indeed, in simulations without cubic terms, one recovers the results from the literature \citep{ishizawa2010turbulence, muraglia2011generation}: direct coupling of the mode $m^*$ with the mode $m^* \pm 1$ drives non-linearly an $m=1$ mode with tearing parity, whose growth rate satisfies $\gamma \approx 2 \gamma^*$. However, such modes never become dynamically important. Only if the unstable modes (with $m \approx m^*$) change parity, will a clearly identifiable magnetic island form in the system, after the non-linear energy transfer across scales takes place. The small-scale islands that form at the unstable scales merge (coalesce) into larger islands over time, allowing the sub-dominant low-$m$ modes, formed by direct coupling, to become dynamically relevant and form large-scale magnetic islands. Thus without the mechanism described here, no large-scale island would be visible in the system (see the supplemetary material for figures illustrating this process).\\
The role of the cubic terms (and partially of $\beta$ and $\partial_x B_{eq}$) can be understood by focusing on the pressure term in Ohm's law: $\frac{\Omega_i \tau_A \rho_*^2}{n} \{ \psi, p_e \}$. In an interchange unstable system, the parity of $\psi_{m^*}$ in the linear (and early non-linear) phase is odd, and non-linear dynamics form modes $m=0$ of pressure and density that also have odd parity (to give profile flattening). The Poisson bracket of two odd functions is odd, multiplied by $n_{m=0}$ leads to an even term, at the same $m^*$ of the linear instability. This changes the parity of the unstable mode.\\
It is also important to remark that the phase change described here happens on much faster time-scales than the resistive time-scale (in the late linear phase it's possible to quantify the average behaviour of the phases described above), and thus is attributed to non-linear (turbulent) behaviour. The non-linear energy transfer also occurs on time-scales faster than the resistive one, so that overall the process is quicker than the resistive time-scale.\\
Summarizing, the novelty of the process described here is the fact that the change in structure of the unstable modes is allowed by a combination of the background parameters and the presence of the cubic terms. Through the above non-linear dynamics, small-scale TDMI form. They allow the otherwise sub-dominant large-scale TDMIs formed by the direct coupling described in the literature \citep{yagi2007nonlinear, muraglia2011generation, dubuit2021turbislands} to appear in the system.\\
These simulations only allowed seeing islands with $m=2$, while one might expect to observe islands with $m=1$: for simulations with strong instability drive, the islands reached the boundaries of the domain, regardless of the size of the latter, before the mode $m=1$ formed, while for simulations with weak drive, the non-linear energy transfer slows down progressively, rendering the computational cost of observing the $m=1$ mode prohibitive. The strong instability case might point to a missing saturation mechanism in our simulations, but this will have to be addressed in future studies.
%Removed -13 words (i.e. added)
% Using the phase information of the field allows to make the analysis independent of the notion of parity, mostly useful in linear systems.

% \section{Interaction of the large scale structures in the non-linear phase}

\begin{figure}
\includegraphics[width = \ColFigWidth, trim=0 0 1cm 3cm, clip]{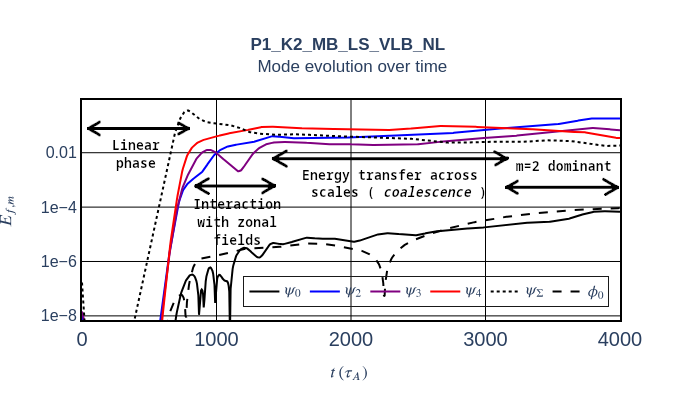}
\caption{\label{fig_history} Evolution of the energies for the simulation with \(\partial_x B_{eq} = 0.01 \) and \( \beta = 1.28 \% \). For the magnetic potential $\psi$ multiple modes are shown to highlight the different phases of the simulation, while for the electrostatic potential $\phi$ only the mode $m=0$ (i.e. the zonal flow) is shown. The curve labeled as \(\psi_\Sigma\) is obtained by summing the energies for all modes with \(10 \leq m \leq 40\).}
\end{figure}

The non-linear dynamics of the system are dominated by the interaction among the structures at the largest scales. These structures are the TDMI (with poloidal mode-number $m=2$), the zonal current (i.e. mode $\psi_{m=0}$) and the zonal flow (i.e. mode $\phi_{m=0}$).\\
The fluctuations of the mode $\psi_0$ form a radial region around the resonance where the poloidal magnetic field and its shear vanish.\\
In the non-linear phase, the flow becomes strongly sheared at the resonance, which is expected since that is where the instability is driven, and also at the radial position where the magnetic island has its maximum radial width. As the island grows in width, this latter region of sheared flow moves with the growing island away from the resonant surface, which is indicative of a strong interdependence between the magnetic island and the localization of the velocity shear.\\
Shown in Fig. \ref{fig_history} is the time evolution of the energies of the (selected) modes of \(\psi\) and the zonal flow  (\(\phi_0\)) for the simulation with \(\partial_x B_{eq} = 0.01 \) and \( \beta = 1.28 \% \). It's visible how in the linear phase the linearly unstable modes (the curve labeled as \(\psi_\Sigma\) is obtained by summing the energies for all modes with \(10 \leq m \leq 40\)) grow and drive, by non-linear coupling, the large-scale modes (\(\psi_2, \psi_3, \psi_4\)). A period of interaction with the zonal fields, visible in the quick oscillations of the zonal current \(\psi_0\), enhances the energy transfer from the small to the large scales. In simulations that did not show the formation of TDMIs this phase is missing, and the zonal fields have much ``smoother'' energy dynamics. Furthermore, in simulations that did not form TDMIs, the large-scale modes remain at low energy levels, while here a shift in the energy distribution is visible quite early in the non-linear phase. The importance of modes $m=0$ on the dynamics of coalescence has already been highlighted also in other contexts such as in \cite{poye2014dynamics}. Late in the simulation the mode \(\psi_2\) becomes dominant, and the large-scale modes acquire significantly larger energies than the turbulent scales. One could expect the mode \(\psi_1\) to become dominant over time, but, as already addressed, this was not observed.\\
To further address the role of the zonal fields, variations on the simulations were run where the fluctuations of the zonal fields were suppressed either through the use of a stronger dissipation only to the mode $m=0$ of the electrostatic potential $\phi$ from the beginning of the simulations, or by completely eliminating fluctuations of the mode $m=0$ of $\psi$ or of $\phi$ from a certain time in the non-linear phase (once the $m=2$ magnetic island was already dominant) onwards. When the zonal flow is suppressed for a simulation that has already reached the non-linear phase, it is observed that the coalescence is halted at the current wave-number, and the island's width stops evolving, indicating that the zonal flow is responsible for the transfer of energy at larger scales (i.e. the coalescence), acting as a catalyst for the formation of TDMIs. This is an unusual behaviour for zonal flows, that usually move energy towards the smaller scales by shearing apart larger eddies \citep{diamond2005zonal}, but it's also peculiar that in these simulations the properties of turbulence aren't much affected by the disappearance of the zonal flow.\\
Suppression of the zonal current for a simulation that has already reached the non-linear phase does the opposite, accelerating the transition of energy to larger scales, while strongly suppressing turbulence. This can be understood as the re-establishing of the background magnetic shear, that limits the region where turbulence can develop. Thus the zonal current acts as an inhibitor to the formation of TDMIs, slowing down the coalescence by creating the low-shear region, that favours the accumulation of energy at the turbulent scales.\\
It has long been known from gyrokinetic analytical calculations \citep{rosenbluth1998poloidal} and simulations \citep{lin1998zonal} that zonal flow saturation is not dominated by fluid dissipation. Given the important role of zonal flows described above, one should check that our results are not reliant on zonal flow dissipation. Starting simulations anew with varied dissipation for the mode $\phi_0$ shows, as expected, the existence of two regimes of zonal flow saturation (ideal and collisional). Specifically, increase of the dissipation of the zonal flow by a factor $50$ doesn’t lead to any effect, while a further factor of $5$ finally gives a factor $\sim 1/5$ on zonal flow saturation levels.  The simulations discussed in this paper are therefore in the ideal regime, where both zonal fields saturate at similar levels regardless of the magnitude of the dissipation.
% Removed -40 words

% \section{Discussion} \label{sec_discussion}
In conclusion, this paper shows how the formation of TDMIs can happen through a process never observed before that involves small-scale islands forming at the unstable scales and then merging to progressively larger scales. This happens through a coalescence process that allows the otherwise negligible large-scale islands formed by direct non-linear coupling to become dominant. Coalescence occurs at a later time than the direct coupling of the most unstable modes reported in the literature (e.g. \cite{muraglia2011generation}), and, since it requires energy transfer across scales, is slower than the direct coupling (direct coupling takes $\approx 10 (\gamma^*)^{-1}$, in Fig. \ref{fig_history} this would be $t \approx 800 \tau_A$, while coalescence takes $\geq 50 (\gamma^*)^{-1}$), but it modifies the dynamics of the system fundamentally, and, in particular, leads to much larger magnetic islands than otherwise observed. Simulations that form magnetic islands, unlike those that don't, undergo a change in the radial structure of the magnetic potential $\psi$, rendering the phase of the mode more uniform over the radial coordinate (Fig. \ref{fig_NL_eigenmode}). The equilibrium magnetic shear of the system, and the presence of the cubic terms in the model, are crucial in allowing this change in parity, and thus the coalescence, to take place. This process is of particular importance for fusion plasmas, since it allows the formation and growth of magnetic islands independent of neoclassical physics, and it might feedback into the dynamics of NTMs. The latter are expected to pose a threat to future devices, operating at higher $\beta$ than existing ones (and than explored here), in particular since the source of the seed islands is, in some cases, unexplained \citep{isayama2013onset}. On the other hand, astrophysical plasmas occur more often in conditions where $\beta \approx 1$, such that pressure fluctuations can play a much larger role on the dynamics of the system, allowing the cubic terms to impact the dynamics much more significantly, making the process described here relevant.\\
The TDMI is capable of driving a strongly sheared poloidal flow at the position where its separatrix has its maximum radial width, and the presence of the zonal flow is needed for the coalescence process to continue throughout the simulation. The zonal flow is mostly driven by the turbulence and the magnetic islands, rather than being regulated by dissipation. Given the strong role in the establishment of Internal Transport Barriers \citep{benkadda2001bursty, ida2018internal} played by zonal fields, this might be an unexpected positive outcome of the presence of TDMIs. The zonal current, instead, slows down the coalescence process, favouring the development of small-scale turbulence by creating a region of weak magnetic shear.\\
% Removed 9 words

\section*{Supplementary Material}
The supplementary material to this letter contains the equations of the model, including normalizations and parameters used to run the simulations. There is also a series of figures illustrating the change of phase and the coalescence through the isocontours of the magnetic potential $\psi$, as well as the crucial role played by the cubic terms, and an explanation of the use of phase information to analyze the parity of the modes.

\begin{acknowledgments}
The authors would like to thank the referees for their constructive and detailed feedback.\\
The authors would like to thank S. Mazzi and M.J. Pueschel for constructive discussions on the topic.\\
The project leading to this publication has received funding from the Excellence Initiative of Aix-Marseille University - A*Midex, a French “Investissements d’Avenir” program AMX-19-IET-013.
The simulations in this article were run thanks to the support of EUROfusion and MARCONI-Fusion.
This work has been carried out within the framework of the EUROfusion Consortium, funded by the European Union via the Euratom Research and Training Programme (Grant Agreement No 101052200 — EUROfusion). Views and opinions expressed are however those of the author(s) only and do not necessarily reflect those of the European Union or the European Commission. Neither the European Union nor the European Commission can be held responsible for them.
This work was partially funded by the Deutsche Forschungsgemeinschaft (DFG, German Research Foundation) under Germany´s Excellence Strategy – EXC 2094 – 390783311.\\

The authors have no conflicts to disclose.\\

The data that support the findings of this study are available from the corresponding author upon reasonable request.
\end{acknowledgments}

\bibliographystyle{unsrt}
\bibliography{./PoP_Letter_open.bib}

\end{document}

\title{Zonal fields as catalysts and inhibitors of turbulence-driven magnetic islands - Supplementary Material}% Force breaks with \\

\author{D. Villa}
 \affiliation{Max Planck Institute for Plasma Physics, Boltzmannstraße 2, 85748, Garching, Germany}%Lines break automatically or can be forced with \\
 \email{daniele.villa@ipp.mpg.de}

\author{N. Dubuit}%
\affiliation{Aix-Marseille Universit\'e, CNRS, PIIM UMR 7345, Marseille, France}

\author{O. Agullo}
\affiliation{Aix-Marseille Universit\'e, CNRS, PIIM UMR 7345, Marseille, France}

\author{X. Garbet}
\affiliation{CEA, IRFM, F-13108 Saint-Paul-Lez-Durance, France}
\affiliation{School of Physical and Mathematical Sciences, Nanyang Technological University, 637371 Singapore}

\date{\today}

\maketitle

\section{Equations of the model}

\begin{widetext}
\begin{eqnarray} \label{eq_psiev_def}
\partial_t \psi &=& \{ \psi, \phi \} - \frac{\Omega_i \tau_A \rho^2_*}{n} \{ \psi, p_e \} + \frac{\eta}{n} \tilde{J_{\|}}
\end{eqnarray}
\begin{eqnarray} \label{eq_omegaev_def}
 \partial_t \mathcal{W} &=& - \{ \phi, \mathcal{W} \} - \tau_i \Omega_i \tau_A \rho^2_* \{ \nabla_{\alpha} \phi , \nabla_{\alpha} p_i \} - u_{\parallel \, i} \{ \psi , \mathcal{W} \} + \{ \psi , J_{\parallel} \} \\ \nonumber
&& + \rho^2_* (\Omega_i \tau_A)^2 \bm{\mathcal{K}_1} \! \cdot \! \nabla(\tau_i p_i + p_e) + \mu \Delta_{\perp} \tilde{\mathcal{W}}
\end{eqnarray}
\begin{eqnarray} \label{eq_piev_def}
\partial_t p_i &=& \! - \frac{5}{3} p_i \{ \psi, u_{\parallel \, i} \} - \{ \phi, p_i \} - u_{\parallel \, i} \{ \psi, p_i \} - \frac{5}{3}p_i \bm{\mathcal{K}_2} \cdot \nabla \phi + \frac{5}{3} \frac{1}{\Omega_i \tau_A} T_i \{ \psi , J_{\parallel} \} \\ \nonumber
&& - \frac{5}{3} \Omega_i \tau_A \rho^2_* \Big( \tau_i p_i \bm{\mathcal{K}_2} \! \cdot \! \nabla T_i  - T_i\bm{\mathcal{K}_2} \! \cdot \! \nabla p_e \Big) + \chi_{\perp \, i} \Delta_{\perp} \tilde{T}_i + \chi_{\parallel \, i} \{ \psi , \{ \psi , \tilde{T}_i \} \}
\end{eqnarray}
\begin{eqnarray} \label{eq_peev_def}
\partial_t p_e &=& - \frac{5}{3} p_e \{ \psi, u_{\parallel \, e} \} - \{ \phi, p_e \} - u_{\parallel \, e} \{ \psi, p_e \} - \frac{5}{3} p_e \bm{\mathcal{K}_2} \! \cdot \! \nabla \phi \\ \nonumber
&& + \frac{5}{3} \Omega_i \tau_A \rho^2_* \bm{\mathcal{K}_2} \! \cdot \! \nabla(T_e p_e) + \chi_{\perp \, e} \Delta_{\perp} \tilde{T}_e + \chi_{\parallel \, e} \{ \psi , \{ \psi , \tilde{T}_e \} \}
\end{eqnarray}
\begin{eqnarray} \label{eq_nev_def}
\partial_t n &=& - \{ \phi, n \} - n \{ \psi, u_{\parallel \, e} \} - u_{\parallel \, e} \{ \psi, n \} -  n \bm{\mathcal{K}_3} \! \cdot \! \nabla \phi + \Omega_i \tau_A \rho^2_* \: \bm{\mathcal{K}_3} \! \cdot \! \nabla p_e + D \Delta_{\perp} \tilde{n}
\end{eqnarray}
\begin{eqnarray} \label{eq_upar_def}
\partial_t u_{\parallel \, i} &=& - \{ \phi , u_{\parallel \, i} \} - \left\{ \psi, \frac{u_{\parallel \, i}^2}{2} \right\} - \frac{(\Omega_i \tau_A)^2 \rho^2_*}{n}  \{ \psi, \tau_i p_i + p_e \} + U_d \Delta_{\perp} \tilde{u}_{\parallel \, i}
\end{eqnarray}
\end{widetext}

where $\psi$ is the poloidal magnetic flux, $\phi$ the electrostatic potential, $p_e$ and $T_e$ the electron pressure and temperature, $p_i$ and $T_i$ the ion pressure and temperature, $n$ the electron density (quasi-neutrality is assumed) and $u_{\parallel \, i/e}$ the ion and electron parallel fluid velocities, with $u_{\parallel \, e} = u_{\parallel \, i} - \frac{J_{\parallel}}{n \, \Omega_i \tau_A}$. The definition of the generalized vorticity is
\begin{equation} \label{eq_omega}
\mathcal{W} = \Delta_{\perp} (\phi + \tau_i \Omega_i \tau_A \rho^2_* p_i)
\end{equation}
meaning that here the Boussinesq approximation is applied, and imposes a condition for the equilibrium profiles, which is always satisfied in simulations unless clearly stated: $\partial_x \phi_{eq} = - \Omega_i \tau_A \rho^2_* \frac{\partial_x p_{i \, eq}}{n_{eq}}$. Applying the Boussinesq approximation means that when evaluating the term \( \nabla \cdot (n \bm{u}_{pol}) \) the density is considered as constant everywhere it appears (including the diamagnetic drift appearing in \(\bm{u}_{pol}\)). The parallel current density is $J_{\parallel} = \Delta_{\perp} \psi$. A ``$\sim$'' above the symbol of the field indicates that only the fluctuating part of the field is retained. $\alpha$ in Eq. \ref{eq_omegaev_def} is an index for the sum over the perpendicular geometrical coordinates, since the explicit expression of the second term on the RHS of Eq. \ref{eq_omegaev_def} is: $ \{ \nabla_\alpha \phi , \nabla_\alpha p_i \} = \{ \partial_x \phi , \partial_x p_i \} + \{ \partial_y \phi , \partial_y p_i \}$. The curvature is $ \bm{\mathcal{K}} = \left( \frac{1}{B} \nabla \times \bm{b} + \nabla \left( \frac{1}{B} \right) \times \bm{b} \right)$, with $\bm{b} = \frac{\bm{B}}{B}$.\\
Note that the model explicitly depends on the electronic beta through the relation $\beta_{e \, 0} = 2 (\Omega_i \tau_A)^2 \rho_*^2$

\begin{table}
\begin{ruledtabular}
\begin{tabular}{c}
$n = n_0 n_N$\\
$B = B_0 B_N$\\
$u_{\parallel} = v_A u_{\parallel \, N}$ \\
$\nabla_{\perp} = L_{\perp}^{-1} \nabla_{\perp \, N}$ \\
$\nabla_{\parallel} X_N = L_{\perp}^{-1} \{ \psi_N, X_N \} - L_z^{-1}\partial_{z \, N} X_N$ \\
$\partial_t = \tau_A^{-1} \partial_{t \, N}$\\
$\psi = B_0 L_{\perp} \psi_N$ \\
$\phi = B_0 L_{\perp} v_A \phi_N$\\
$\tau_i = T_i / T_e$ $\qquad \qquad$ $T_0 = T_{0 \, e} + T_{0 \, i} = \frac{T_0}{1+\tau_i} + \frac{\tau_i T_0}{1+\tau_i}$\\
$\Omega_i = eB_0/m_i$\\
$\rho^2_* =\frac{T_{0 \, e}}{m_i} \frac{1}{\Omega^2_i}  \frac{1}{L^2_{\perp}}$ = $\frac{T_0}{m_i} \frac{1}{\Omega^2_i}  \frac{1}{L^2_{\perp}} \frac{1}{1+\tau_i} = \frac{\beta_{e\,0}}{2n_0}$\\
$\beta_{e \, 0} = 2 (\Omega_i \tau_A)^2 \rho_*^2 $\\
$p_e = \frac{p_0}{(1 + \tau_i) \rho^2_* \Omega_i \tau_A } \, p_{e \, N} $\\
$p_i = \frac{p_0 \, \tau_i}{(1 + \tau_i)\rho^2_* \Omega_i \tau_A} \, p_{i \, N} $\\
$\bm{\mathcal{K}_3} = \bm{\mathcal{K}_2} = -(5/3) \rho^2_* \bm{\mathcal{K}_1}$\\
\end{tabular}
\caption{\label{table_normalization} Normalization used for the equations of the model. Quantities with subscript ``$0$'' are reference values at the resonant surface. $v_A = \frac{c B_0}{\sqrt{m n_0}}$ is the Alfven velocity at resonant surface. $L_{\perp}$ is a characteristic perpendicular length of the system. $\bm{\mathcal{K}_3}$ and $\bm{\mathcal{K}_2}$ are paremeters introduced to manipulate linear spectrum.}
\end{ruledtabular}
\end{table}

\begin{table}
\begin{ruledtabular}
\begin{tabular}{ccccc}
$\eta$ & $1 \cdot 10^{-5}$ &\qquad \qquad & $\rho_*^2$ & $1.66 \cdot 10^{-4}$ \\
$\mu$ & $5 \cdot 10^{-4}$ &\qquad \qquad & $\mathcal{K}_1$ & $-0.5 \; [-1]$  \\
$\chi_{\perp \, e/i}$ & $5 \cdot 10^{-5}$ &\qquad \qquad & $\mathcal{K}_2 = \mathcal{K}_3$ & $-3 \cdot 10^{-1}$ \\
$\chi_{\parallel \, e}$ & $5.0\cdot10^{2}$ &\qquad \qquad & $L_y$ & $6 \pi$ \\
$\chi_{\parallel \, i}$ & $8.0 \cdot 10^{0}$ &\qquad \qquad & $\tau_i = T_i/T_e$ & $1.0$ \\
$D$ & $1 \cdot 10^{-4}$ & \qquad \qquad & $\Omega_i \tau_A$ & $3.12 \; [6.24]$ \\
$U_d$ & $5 \cdot 10^{-5}$ & \qquad \qquad & $L_x$ & $10$\\
\end{tabular}
\caption{\label{tab_typical_parameters_2D} Dissipative and dimensionless parameters used for the 2D non-linear simulations in slab geometry. The value of $\Omega_i \tau_A$ in brackets is for the higher-$\beta$ case. The value of $\mathcal{K}_1$ in brackets is for the stronger drive case.}
\end{ruledtabular}
\end{table}

\section{Parity analysis through complex argument differences}
\begin{figure}[H]
\includegraphics[width=.49\textwidth, trim=0 0 0 3cm, clip]{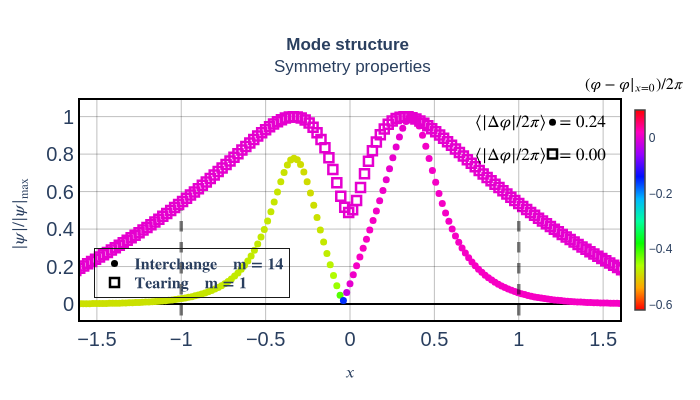}
\caption{\label{fig_puremode}Comparison of the linear modes for two different simulations: one interchange unstable (shown with circles), the other tearing unstable (shown with empty squares). The asymmetry in the absolute value of the interchange unstable mode is due to the cubic terms.}
\end{figure}

Since the paper employs an unconventional way to define the parity of the modes, using the average argument of the complex-valued functions, rather than comparing values across a center of symmetry, here is provided some more insight on this point. Fig. \ref{fig_puremode} shows the eigenfunctions of two different simulations in the linear phase: one interchange unstable (shown with circles), the other tearing unstable (shown with empty squares). The left-axis shows the normalized asbolute value of the mode, while the colorscale shows the difference between the phase of the function at any given radial position to the phase of the function at a chosen radial position (in this case the $x=0$ position). Note that the slight asymmetry in the absolute value of the interchange unstable mode is an effect due to the cubic terms when the equilibrium pressure profile is not flat (in the tearing unstable simulation this profile was flat), an effect that will be investigated in a paper currently in preparation. As evident, an even mode (i.e. an unstable tearing mode) has constant phase across the resonant position, whereas the interchange unstable mode has a phase change of $\pi/2$ across the resonant position. In these linear simulations, these phase differences will be exact, as is evidenced by the average values $<\vert \Delta \varphi \vert/2\pi>$ computed and shown in Fig. \ref{fig_puremode}.In the non-linear simulations of the paper, these averaged values will not be exact as is the case in this linear example, but lower average phase differences indicate a more relevant tearing parity of the mode, whereas higher values indicate dominant interchange parity (i.e. that parity of the mode changes more across the radius, as is the case in the interchange parity simulations).

\section{The coalescence process}
\begin{figure}[H]
\includegraphics[width=.45\textwidth, trim=0 0 0 3cm, clip]{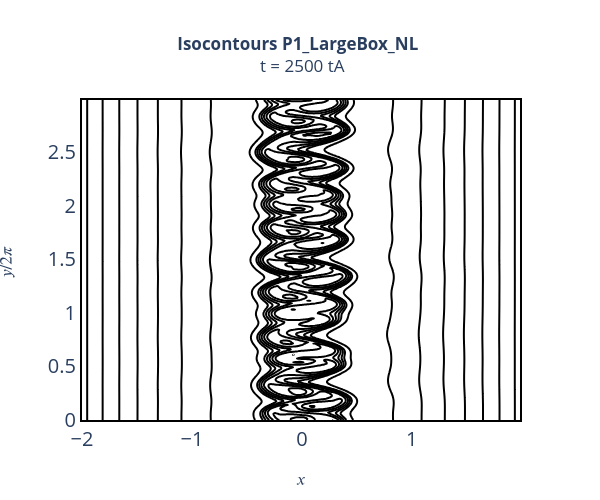}
\caption{\label{fig_nokoala}Isocontours of $\psi$ for a simulation with $\gamma^* = 0.015$ and $\partial_x B_{eq} = 0.02$, that will \textbf{not} undergo coalescence, showing interchange parity in the linear phase.}
\end{figure}

\begin{figure}[H]
\includegraphics[width=.45\textwidth, trim=0 0 0 3cm, clip]{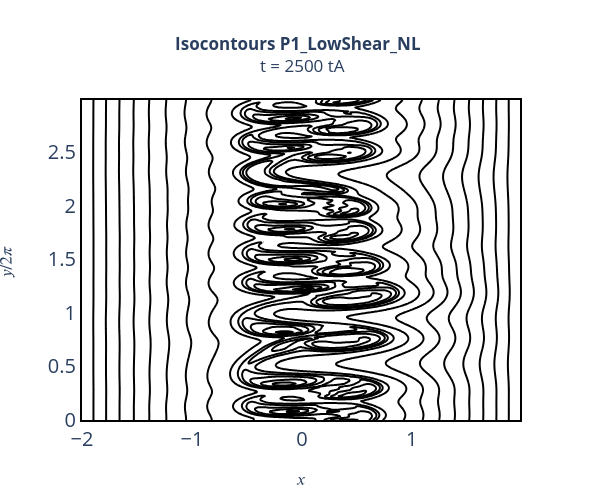}
\caption{\label{fig_koala}Isocontours of $\psi$ for a simulation with $\gamma^* = 0.018$ and $\partial_x B_{eq} = 0.01$, that will undergo coalescence, showing interchange parity in the linear phase.}
\end{figure}

\begin{figure}[H]
\includegraphics[width=.45\textwidth, trim=0 0 0 3cm, clip]{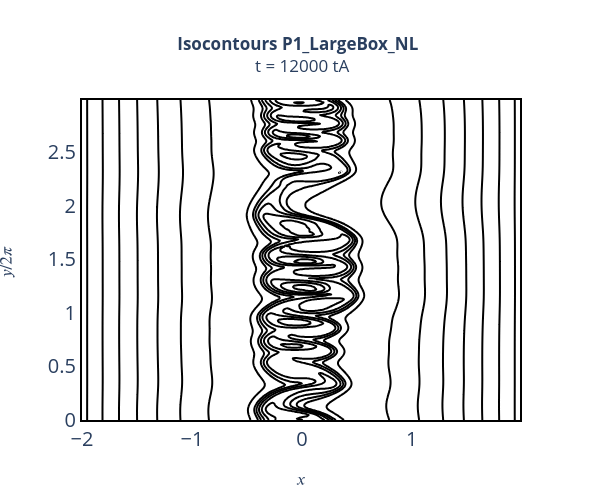}
\caption{\label{fig_nokoala_late}Isocontours of $\psi$ for a simulation with $\gamma^* = 0.015$ and $\partial_x B_{eq} = 0.02$, that will \textbf{not} undergo coalescence, showing interchange parity in the early non-linear phase.}
\end{figure}

\begin{figure}[H]
\includegraphics[width=.45\textwidth, trim=0 0 0 3cm, clip]{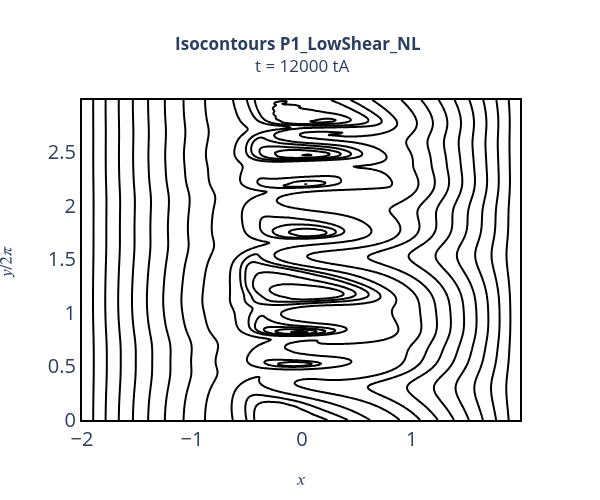}
\caption{\label{fig_koala_late}Isocontours of $\psi$ for a simulation with $\gamma^* = 0.018$ and $\partial_x B_{eq} = 0.01$, that will undergo coalescence, showing tearing-like parity in the early non-linear phase.}
\end{figure}

\begin{figure}[H]
\includegraphics[width=.45\textwidth, trim=0 0 0 3cm, clip]{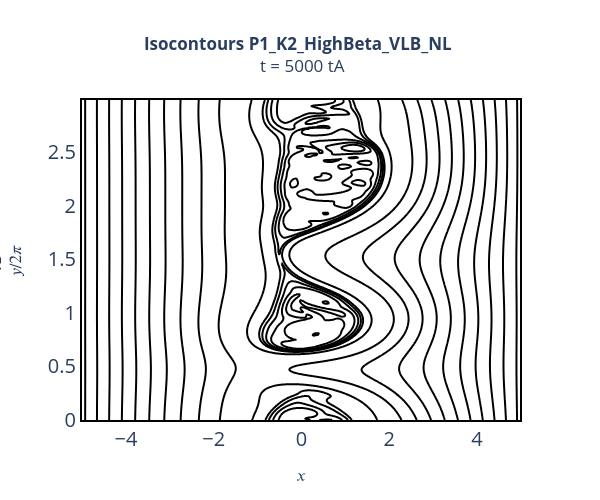}
\caption{\label{fig_late}Isocontours of $\psi$ in the late non-linear phase of a simulation with $\gamma^* = 0.083$ and $\partial_x B_{eq} = 0.02$ that shows coalescence. Notice the X-points at $y \approx 3 \pi$ and $y \approx \pi$.}
\end{figure}

\begin{figure}[H]
\includegraphics[width=.45\textwidth, trim=0 0 0 3cm, clip]{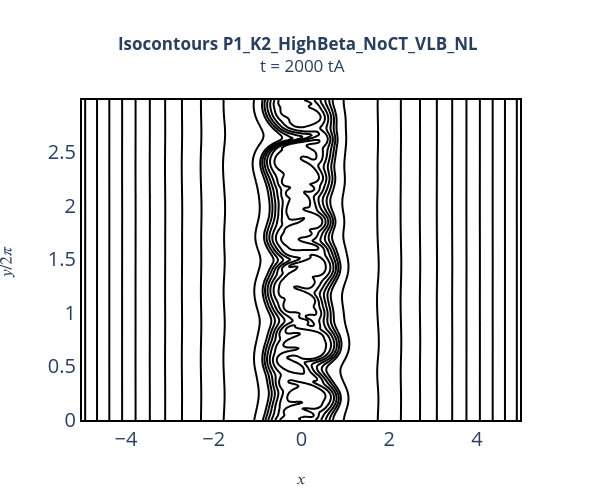}
\caption{\label{fig_late_noct}Isocontours of $\psi$ in the late non-linear phase of a simulation with $\gamma^* = 0.083$ and $\partial_x B_{eq} = 0.02$ where \textbf{the cubic terms were not included in the dynamics} (compare to Fig. \ref{fig_late}). Notice that the modes retain interchange parity and remain limited in width.}
\end{figure}

\begin{figure}[H]
\includegraphics[width=.45\textwidth, trim=0 0 0 3cm, clip]{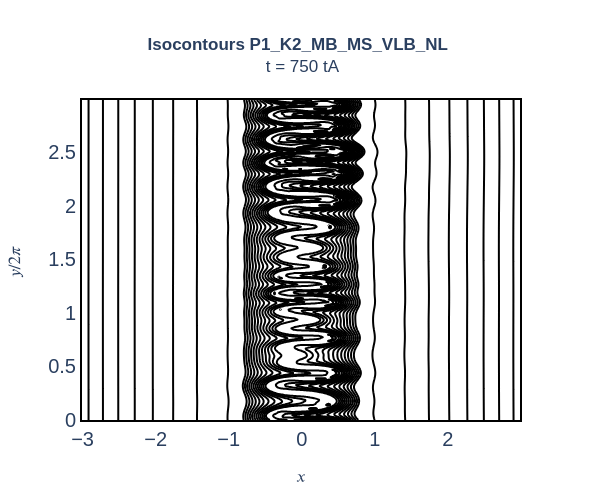}
\caption{\label{fig_c1}Isocontours of $\psi$ in the late linear phase for the simulation with $\gamma^* = 0.049$ and $\partial_x B_{eq} = 0.02$.}
\end{figure}

\begin{figure}[H]
\includegraphics[width=.45\textwidth, trim=0 0 0 3cm, clip]{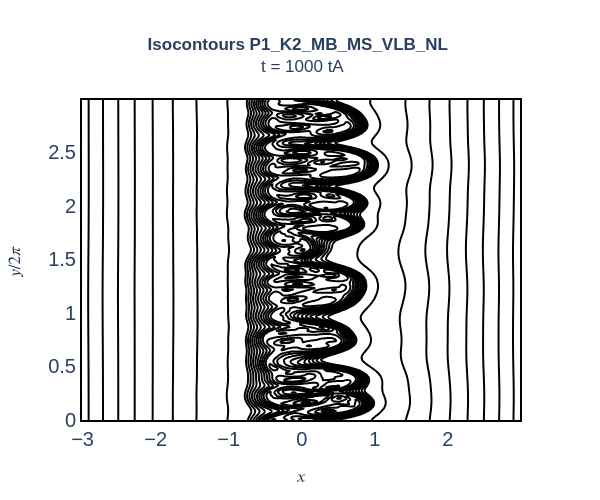}
\caption{\label{fig_c2}Isocontours of $\psi$ in the early non-linear phase for the simulation with $\gamma^* = 0.049$ and $\partial_x B_{eq} = 0.02$.}
\end{figure}

\begin{figure}[H]
\includegraphics[width=.45\textwidth, trim=0 0 0 3cm, clip]{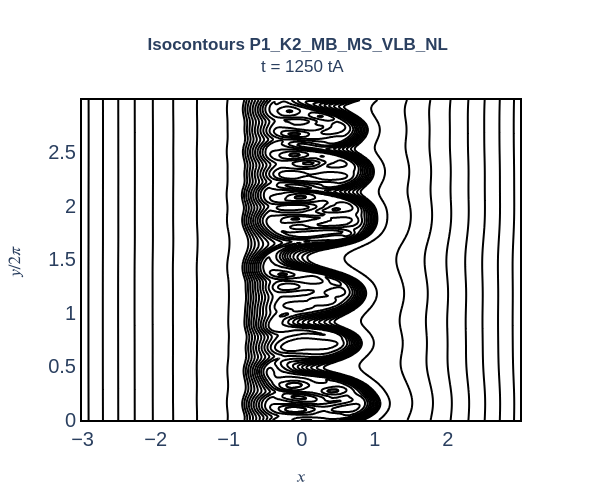}
\caption{\label{fig_c3}Isocontours of $\psi$ in the fully developed non-linear phase for the simulation with $\gamma^* = 0.049$ and $\partial_x B_{eq} = 0.02$.}
\end{figure}

\begin{figure}[H]
\includegraphics[width=.45\textwidth, trim=0 0 0 3cm, clip]{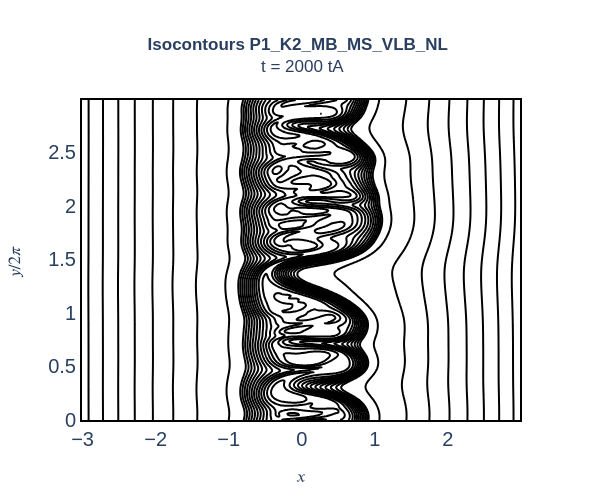}
\caption{\label{fig_c4}Isocontours of $\psi$ in the late non-linear phase for the simulation with $\gamma^* = 0.049$ and $\partial_x B_{eq} = 0.02$.}
\end{figure}

\begin{figure}[H]
\includegraphics[width=.45\textwidth, trim=0 0 0 3cm, clip]{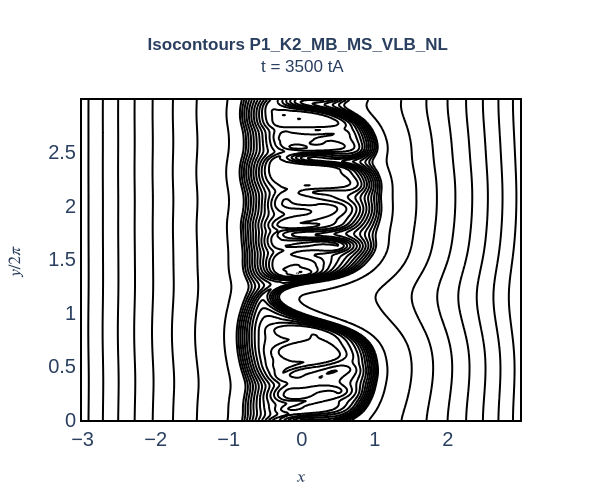}
\caption{\label{fig_c5}Isocontours of $\psi$ at the latest point available in the non-linear phase for the simulation with $\gamma^* = 0.049$ and $\partial_x B_{eq} = 0.02$.}
\end{figure}

\begin{figure}[H]
\includegraphics[width=.45\textwidth, trim=0 0 0 3cm, clip]{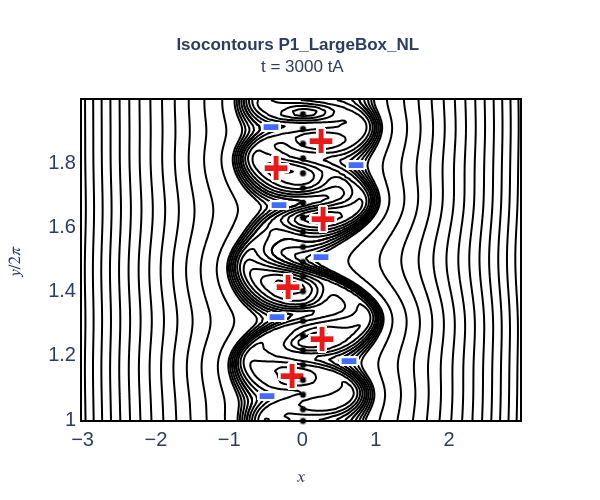}
\caption{\label{fig_c6}Zoomed in detail of the isocontours of $\psi$ for a simulation with $\gamma^* = 0.015$ and $\partial_x B_{eq} = 0.02$, that does \textbf{not} undergo coalescence, showing that the fluctuations of $\psi$ (in most cases) change sign across the resonant position, maintaining a dominant interchange parity.}
\end{figure}

\begin{figure}[H]
\includegraphics[width=.45\textwidth, trim=0 0 0 3cm, clip]{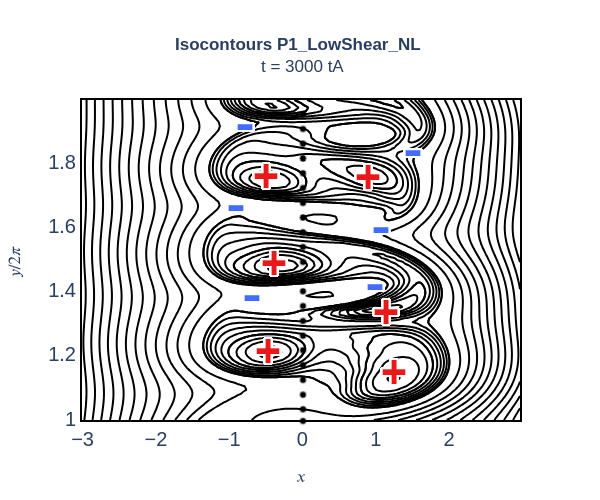}
\caption{\label{fig_c7}Zoomed in detail of the isocontours of $\psi$ for a simulation with $\gamma^* = 0.018$ and $\partial_x B_{eq} = 0.01$, as it is undergoing coalescence, showing how the fluctuations of $\psi$ of the same sign tend to align horizontally, changing the phase of the mode.}
\end{figure}

Here are shown some figures to illustrate the difference between simulations that show coalescence (weaker magnetic shear $\partial_x B_{eq} = 0.01$ and slightly stronger instability drive $\gamma^* = 0.018$) and those that don't (stronger magnetic shear $\partial_x B_{eq} = 0.02$ and slightly weaker instability drive $\gamma^* = 0.015$). These simulations were not directly used in the paper, but clearly illustrate the different behaviours of the two regimes, while also having very similar instability drives, thus highlighting the role of the shear in the dynamics. Using the isocontours of the magnetic potential $\psi$ one can see how in the linear phase (Figs. \ref{fig_nokoala} and \ref{fig_koala}) both simulations are unstable to interchange modes, with the simulation with weaker magnetic shear showing radially broader modes. In the early non-linear phase, the simulation with stronger magnetic shear (Fig. \ref{fig_nokoala_late}) retains dominant interchange parity in its modes, whereas the simulation with weaker magnetic shear (Fig. \ref{fig_koala_late}) has modes with noticeably more tearing-like parity. Fig. \ref{fig_late} shows the late non-linear phase of a simulation that undergoes coalescence (this one part of the set used in the paper), with turbulent regions enclosed within the magnetic islands that have formed non-linearly. Fig. \ref{fig_late_noct} shows the late non-linear phase of a simulation run with the same parameters as the simulation giving Fig. \ref{fig_late} but without including the cubic terms in the dynamics, showing their crucial role in allowing the mechanism described in the paper to take place. Figs. \ref{fig_c1} - \ref{fig_c5} show a sequence of isocontours for the simulation with $\gamma^* = 0.049$ and $\partial_x B_{eq} = 0.02$ (zoomed in to the region $x \in [-3,3]$) to more clearly illustrate the merging of structures from small to large scales. Figs. \ref{fig_c6} - \ref{fig_c7} show a zoomed-in detail of the isocontours for two simulations in the non-linear phase, where the one undergoing coalescence can be seen having regions with fluctuations of $\psi$ of the same sign aligning horizontally, effectively changing the parity of the mode.\\
Note: $\gamma^*$ is the growth rate of the most unstable mode in the linear phase.

